\documentclass[12pt]{article}
\usepackage{amssymb,amsmath,epsfig}

\begin{document}
\title{\bf Energy Conditions Constraints and Stability of
Power Law Solutions in $f(R,T)$ Gravity}
\author{Muhammad SHARIF \thanks{msharif.math@pu.edu.pk} and Muhammad
ZUBAIR \thanks{mzubairkk@gmail.com}\\
Department of Mathematics, University of the Punjab,\\
Quaid-e-Azam Campus, Lahore-54590, Pakistan.}

\date{}

\maketitle

\begin{abstract}
The energy conditions are derived in the context of $f(R,T)$
gravity, where $R$ is the Ricci scalar and $T$ is the trace of the
energy-momentum tensor, which can reduce to the well-known
conditions in $f(R)$ gravity and general relativity. We present the
general inequalities set by the energy conditions in terms of
Hubble, deceleration, jerk and snap parameters. In this study, we
concentrate on two particular models of $f(R,T)$ gravity namely,
$f(R)+\lambda{T}$ and $R+2f(T)$. The exact power-law solutions are
obtained for these two cases in homogeneous and isotropic $f(R,T)$
cosmology. Finally, we find certain constraints which have to be
satisfied to ensure that power law solutions may be stable and match
the bounds prescribed by the energy conditions.
\end{abstract}
{\bf Keywords:} $f(R,T)$ gravity; Raychaudhuri equation; Energy
conditions; Power law.\\
{\bf PACS:} 04.50.-h; 04.50.Kd; 98.80.Jk; 98.80.Cq.

\section{Introduction}

Recent astrophysical observations form supernova type Ia$^{{1})}$,
cosmic microwave background anisotropies$^{{2})}$, large scale
structure$^{{3})}$, baryon acoustic oscillations$^{{4})}$ and weak
lensing$^{{5})}$ indicate that the  universe is accelerating in the
current epoch. The most promising feature of the universe is the
dominance of exotic energy component with large negative pressure,
known as \emph{dark energy} (DE). A number of alternative models
have been proposed in the framework of general relativity (GR) to
explain the role of DE in the present cosmic acceleration.
Unfortunately, up to now, no suitable candidate is found, which
boosts our interest in modified theories of gravity. Firstly, the
Einstein-Hilbert action has been modified by replacing scalar
curvature $R$ by an arbitrary function of $R$, this theory is known
as $f(R)$ gravity$^{{6})}$. The other alternative theories of
gravity include $f(\mathcal{T})$ gravity$^{{7})}$, where
"$\mathcal{T}$" is the torsion scalar in teleparallel gravity,
Gauss-Bonnet gravity$^{{8})}$ and $f(R,T)$ gravity$^{{9})}$.

The $f(R,T)$ gravity is the generalization of $f(R)$ gravity
involving the dependence of the trace of energy-momentum tensor $T$.
The dependence of $T$ may be induced by exotic imperfect fluids or
quantum effects. The cosmological reconstruction of $f(R,T)$ gravity
has been studied in recent literature$^{{9-13})}$. In a
paper$^{{9})}$, the reconstruction of FRW cosmology is presented for
$f(R,T)=R+2f(T)$ model. Houndjo and Piattella$^{{10})}$ constructed
$f(R,T)$ models describing the unification as well as transition of
matter dominated phase to late accelerating phase. The chaplygin gas
$f(R,T)$ models are investigated in$^{{11-12})}$ and it is shown
that dust fluid reproduces $\Lambda$CDM, Einstein static universe
and phantom cosmology$^{{12})}$. In our previous work$^{{13})}$, we
have reconstructed some explicit models of $f(R,T)$ gravity for
anisotropic universe and explored the phantom era of dark energy. We
have also discussed the validity of first and second laws of
thermodynamics in this modified gravity$^{{14})}$. The existence of
exact power law solutions for FRW spacetime has been investigated in
modified theories of gravity$^{{15-16})}$. Here, we shall show that
FRW power law solutions exist for a particular class of $f(R,T)$
gravity.

The classical energy conditions of GR are profound to the
Hawking-Penrose singularity theorems and classical black hole laws
of thermodynamics$^{{17})}$. These conditions have been used to
address several important issues in GR and cosmology$^{{18})}$. Many
authors have investigated the energy conditions in the context of
modified theories including $f(R)$ gravity$^{{19-20})}$, $f(R)$
gravity with nonminimal coupling to matter$^{{21})}$, modified
Gauss-Bonnet gravity$^{{22})}$, modified $f(G)$ gravity with
curvature-matter coupling$^{{23})}$, Brans-Dicke theory$^{{24})}$,
$f(\mathcal{T})$ gravity$^{{25})}$ and $f(R,T)$ gravity$^{{26})}$.
In recent work$^{{26})}$, the authors study the energy conditions
for a special form of $f(R,T)$ gravity, $f(R,T)=R+2f(T)$ and
discussed the stability of two $f(T)$ models.

In this work, we are interested to set energy conditions bounds on
exact power law solutions in $f(R,T)$ gravity. The FRW power law
solutions are obtained for $f(R,T)=f(R)+{\lambda}T$ and
$f(R,T)=R+2f(T)$ gravity. We derive the energy conditions for more
general as well as particular class of $f(R,T)$ gravity. The
standard form of energy conditions in GR and $f(R)$ gravity can be
recovered in the limit of $f(R,T)=f(R)$ and $f(R,T)=R$. We show that
for $f(R,T)=f(R)+{\lambda}T$, the null energy condition (NEC) and
strong energy condition (SEC) can be derived by using the
Raychaudhuri equation with the requirement that gravity is
attractive. The resulting inequalities for NEC and SEC are
equivalent to the energy conditions obtained in terms of effective
energy-momentum tensor.

The paper is organized as follows: In next section, we present the
general formulation of the field equations of $f(R,T)$ gravity in
FRW cosmology. In section \textbf{3}, the energy conditions are
derived and hence presented in terms of deceleration $(q)$, jerk
$(j)$ and snap $(s)$ parameters. Section \textbf{4} is devoted to
obtain exact power law solutions for two specific forms of $f(R,T)$
gravity. We also analyze the constraints of energy conditions for
these models. In section \textbf{5}, we investigate the perturbation
and stability of power law solutions. Finally, section \textbf{6}
summarizes the obtained results.

\section{$f(R,T)$ Gravity}

The $f(R,T)$ theory of gravity is an interesting modification to the
Einstein gravity by introducing an arbitrary function of scalar
curvature $R$ and trace of the energy-momentum tensor $T$. The
action for this theory coupled with matter Lagrangian
$\mathcal{L}_{(matter)}$ is given by$^{{9})}$
\begin{equation}\label{1}
\mathcal{A}=\int{dx^4\sqrt{-g}\left[f(R,T)+\mathcal{L}_{(matter)}\right]},
\end{equation}
where $g$ is the determinant of the metric tensor $g_{\mu\nu}$, we
use the units $8\pi{G}=c=1$. The energy-momentum tensor of matter is
defined as$^{{27})}$
\begin{equation}\label{2}
T_{{\mu}{\nu}}=-\frac{2}{\sqrt{-g}}\frac{\delta(\sqrt{-g}
{\mathcal{\mathcal{L}}_{m}})}{\delta{g^{{\mu}{\nu}}}}.
\end{equation}
Varying this action with respect to the metric tensor, we obtain the
field equations of $f(R,T)$ gravity as
\begin{eqnarray}\label{3}
&&R_{{\mu}{\nu}}f_{R}(R,T)-\frac{1}{2}g_{{\mu}{\nu}}f(R,T)+(g_{{\mu}{\nu}}
{\Box}-{\nabla}_{\mu}{\nabla}_{\nu})f_{R}(R,T)\nonumber\\&=&
T_{{\mu}{\nu}}-f_{T}(R,T)T_{{\mu}{\nu}}-f_{T}(R,T)\Theta_{{\mu}{\nu}},
\end{eqnarray}
where $f_{R}(R,T)$ and $f_{T}(R,T)$ denote derivatives of $f(R,T)$
with respect to $R$ and $T$ respectively;
${\Box}=g^{\mu\nu}{\nabla}_{\mu}{\nabla}_{\mu}$ is the d'Alembert
operator, ${\nabla}_{\mu}$ is the covariant derivative associated
with the Levi-Civita connection of the metric tensor and
$\Theta_{{\mu}{\nu}}$ is defined by
\begin{equation}\label{4}
\Theta_{{\mu}{\nu}}=\frac{g^{\alpha{\beta}}{\delta}T_{{\alpha}{\beta}}}
{{\delta}g^{\mu{\nu}}}=-2T_{{\mu}{\nu}}+g_{\mu\nu}\mathcal{L}_m
-2g^{\alpha\beta}\frac{\partial^2\mathcal{L}_m}{{\partial}
g^{\mu\nu}{\partial}g^{\alpha\beta}}.
\end{equation}

The contribution to the energy momentum tensor of matter is defined
as
\begin{equation*}
T_{{\mu}{\nu}}=({\rho}+p)u_{\mu}u_{\nu}-pg_{{\mu}{\nu}},
\end{equation*}
where $u_{\mu}$ is the four velocity of the fluid, $\rho$ and $p$
denote the energy density and pressure, respectively. We can take
$\mathcal{L}_{(matter)}=-p$, then $\Theta_{{\mu}{\nu}}$ becomes
\begin{equation}\label{5}
\Theta_{{\mu}{\nu}}=-2T_{{\mu}{\nu}}-pg_{{\mu}{\nu}}.
\end{equation}
Consequently, the field equations (\ref{3}) can be expressed as
effective Einstein field equations of the form
\begin{equation}\label{6}
R_{{\mu}{\nu}}-\frac{1}{2}Rg_{{\mu}{\nu}} =T_{{\mu}{\nu}}^{eff},
\end{equation}
where $T_{{\mu}{\nu}}^{eff}$ is the effective energy-momentum tensor
in $f(R,T)$ gravity which is defined as
\begin{eqnarray}\nonumber
{T}_{{\mu}{\nu}}^{eff}&=&\frac{1}{f_{R}(R,T)}\left[(1+f_T(R,T))T_{\mu\nu}
+pg_{\mu\nu}f_T(R,T)+\frac{1}{2}(f(R,T)\right.\\\label{7}&-&\left.Rf_{R}(R,T))g_{\mu\nu}+
({\nabla}_{\mu}{\nabla}_{\nu}-g_{{\mu}{\nu}}{\Box})f_{R}(R,T)\right].
\end{eqnarray}

We consider the homogeneous and isotropic flat FRW spacetime as
\begin{equation}\label{8}
ds^{2}=dt^2-a^2(t)d\textbf{x}^2,
\end{equation}
where $a(t)$ is the scale factor and $d\textbf{x}^2$ contains the
spatial part of the metric. In the FRW background, the field
equations may be rewritten as
\begin{equation}\label{9}
3H^2=\rho_{eff}, \quad -(2\dot{H}+3H^2)=p_{eff},
\end{equation}
where $\rho_{eff}$ and $p_{eff}$ are the energy density and pressure
respectively, defined as
\begin{eqnarray}\label{10}
{\rho}_{eff}&=&\frac{1}{f_R}\left[\rho+(\rho+p)f_T+\frac{1}{2}
(f-Rf_{R})-3H(\dot{R}f_{RR}+\dot{T}f_{RT})\right],\\\nonumber
{p}_{eff}&=&\frac{1}{f_R}\left[p-\frac{1}{2}
(f-Rf_{R})+(\ddot{R}+2\dot{R}H)f_{RR}+\dot{R}^2f_{RRR}\right.\\\label{11}&+&\left.
2\dot{R}\dot{T}f_{RRT}+(\ddot{T}+2\dot{T}H)f_{RT}+\dot{T}^2f_{RTT}\right],
\end{eqnarray}
the Hubble parameter $H$ is defined by $H=\dot{a}/a$ and dot denotes
derivative with respect to cosmic time $t$.

\section{Energy Conditions}

Raychaudhuri equation is the key to SEC and NEC together with the
requirement that gravity is attractive for a spacetime manifold
endowed with a metric $g_{\mu\nu}$$^{{17})}$. For the congruence of
timelike geodesics defined by vector field $u^\mu$, the Raychaudhuri
equation reads,
\begin{equation}\label{12}
\frac{d{\theta}}{d\tau}=-\frac{1}{3}{\theta}^2-{\sigma}^{\mu\nu}{\sigma}_{\mu\nu}
+{\omega}^{\mu\nu}{\omega}_{\mu\nu}-R_{\mu\nu}u^{\mu}u^{\nu},
\end{equation}
where $R_{\mu\nu}$ is the Ricci tensor, and $\theta$,
${\sigma}_{\mu\nu}$ and ${\omega}_{\mu\nu}$ are the expansion
parameter, the shear and the rotation associated with the congruence
respectively. The evolution equation for $\theta$, the expansion
scalar of a congruence of null geodesics defined by the null vector
field $\kappa^\mu$, is given by
\begin{equation}\label{13}
\frac{d{\theta}}{d\tau}=-\frac{1}{2}{\theta}^2-{\sigma}^{\mu\nu}{\sigma}_{\mu\nu}
+{\omega}^{\mu\nu}{\omega}_{\mu\nu}-R_{\mu\nu}\kappa^{\mu}\kappa^{\nu}.
\end{equation}
Raychaudhuri equation is known to be purely geometric and hence,
develops no reference to any theory of gravity. As the shear tensor
is purely spatial ${\sigma}^{\mu\nu}{\sigma}_{\mu\nu}\geqslant0$,
thus, for any hypersurface of orthogonal congruence
$(\omega_{\mu\nu}=0)$, the conditions for attractive gravity become
\begin{equation}\label{14}
\textbf{SEC}: \quad R_{\mu\nu}u^{\mu}u^{\nu}\geqslant0, \quad
\textbf{NEC}: \quad R_{\mu\nu}\kappa^{\mu}\kappa^{\nu}\geqslant0.
\end{equation}

One can use the field equations of any gravity to relate
$R_{\mu\nu}$ to the energy-momentum tensor $T_{\mu\nu}$. Thus, the
combination of the field equations and Raychaudhuri equations can
set the physical conditions for the energy-momentum tensor. In the
framework of GR, the conditions (\ref{14}) can be written as
\begin{equation}\label{15}
R_{\mu\nu}u^{\mu}u^{\nu}=(T_{\mu\nu}-\frac{T}{2}g_{\mu\nu})
u^{\mu}u^{\nu}\geqslant0, \quad
R_{\mu\nu}\kappa^{\mu}\kappa^{\nu}=T_{\mu\nu}\kappa^{\mu}
\kappa^{\nu}\geqslant0.
\end{equation}
For perfect fluid, this equation is reduced to the well-known form
of SEC and NEC in GR,
\begin{equation}\label{16}
\rho+3p\geqslant0, \quad \quad \rho+p\geqslant0.
\end{equation}
In modified theories of gravity including $f(R)$ and
$f(\mathcal{T})$ gravity, $R_{\mu\nu}$ can be obtained in terms of
the energy-momentum tensor by using the corresponding field
equations. However, this does not seem apparent in $f(R,T)$ gravity.

We consider the effective energy-momentum tensor $T_{\mu\nu}^{eff}$,
so that the conditions in Raychaudhuri equations are represented as
\begin{equation*}
(T_{\mu\nu}^{eff}-\frac{T^{eff}}{2}g_{\mu\nu})
u^{\mu}u^{\nu}\geqslant0 \quad \&  \quad
T_{\mu\nu}^{eff}\kappa^{\mu} \kappa^{\nu}\geqslant0.
\end{equation*}
Hence, the energy conditions in GR can be applied by replacing
energy density $\rho$ and pressure $p$ by $\rho_{eff}$ and
$p_{eff}$, respectively. Since the Raychaudhuri equation holds for
any geometrical theory of gravity, we will keep the physical
motivation of focussing of geodesic congruences along with
attractive property of gravity to develop the energy conditions in
$f(R,T)$ gravity. We also assume that standard matter obey the four
energy conditions. Using the effective modified field equations
(\ref{6}), the energy conditions for $f(R,T)$ gravity are given by
\begin{eqnarray}\nonumber
\textbf{NEC}:\\\nonumber
{\rho}_{eff}+p_{eff}&=&\frac{1}{f_R}\left[(\rho+p)(1+f_T)+(\ddot{R}-\dot{R}H)f_{RR}
+\dot{R}^2f_{RRR}\right.\\\label{17}&+&\left.2\dot{R}\dot{T}f_{RRT}
+(\ddot{T}-\dot{T}H)f_{RT}+\dot{T}^2f_{RTT}\right]\geqslant0,\\\nonumber
\textbf{WEC}:\\\nonumber
{\rho}_{eff}&=&\frac{1}{f_R}\left[\rho+(\rho+p)f_T+\frac{1}{2}
(f-Rf_{R})-3H(\dot{R}f_{RR}\right.\\\label{18}&+&\left.\dot{T}f_{RT})\right]\geqslant0,
\quad {\rho}_{eff}+p_{eff}\geqslant0,\\\nonumber
\end{eqnarray}
\begin{eqnarray}\nonumber
\textbf{SEC}:\\\nonumber
{\rho}_{eff}+3p_{eff}&=&\frac{1}{f_R}\left[(\rho+3p)+(\rho+p)f_T-f+Rf_{R}
+3\dot{R}^2f_{RRR}\right.\\\nonumber&+&\left.3(\ddot{R}+\dot{R}H)f_{RR}+6\dot{R}\dot{T}f_{RRT}
+3(\ddot{T}+\dot{T}H)f_{RT}\right.\\\label{19}&+&\left.3\dot{T}^2f_{RTT}\right]\geqslant0,
\quad {\rho}_{eff}+p_{eff}\geqslant0,\\\nonumber
\textbf{DEC}:\\\nonumber
{\rho}_{eff}-p_{eff}&=&\frac{1}{f_R}\left[(\rho-p)+(\rho+p)f_T+f-Rf_{R}
-\dot{R}^2f_{RRR}\right.\\\nonumber&-&\left.(\ddot{R}+5\dot{R}H)f_{RR}-2\dot{R}\dot{T}f_{RRT}
-(\ddot{T}+5\dot{T}H)f_{RT}\right.\\\label{20}&-&\left.\dot{T}^2f_{RTT}\right]\geqslant0,
\quad {\rho}_{eff}+p_{eff}\geqslant0, \quad {\rho}_{eff}\geqslant0.
\end{eqnarray}
The inequalities (\ref{17})-(\ref{20}) represent the null, weak,
strong and dominant energy conditions in the context of $f(R,T)$
gravity for FRW spacetime.

We define the Ricci scalar and its derivatives in terms of
deceleration, jerk and snap parameters as
\begin{eqnarray}\nonumber
&&R=-6H^2(1-q), \quad \dot{R}=-6H^3(j-q-2), \\\label{21}
&&\ddot{R}=-6H^4(s+q^2+8q+6),
\end{eqnarray}
where$^{{28})}$
\begin{eqnarray}\nonumber
q=-\frac{1}{H^2}\frac{\ddot{a}}{a}, \quad
j=\frac{1}{H^3}\frac{\dddot{a}}{a}, \quad and \quad
s=\frac{1}{H^4}\frac{\ddddot{a}}{a}.
\end{eqnarray}
Using the above definitions, the energy conditions
(\ref{17})-(\ref{20}) can be rewritten as
\begin{eqnarray}\nonumber
\textbf{NEC}&:& (\rho+p)(1+f_T)-6H^4(s-j
+(q+1)(q+8))f_{RR}+H^4[6H(j\\\nonumber&-&q-2)]^2
f_{RRR}-12H^3(j-q-2)\dot{T}f_{RRT}+(\ddot{T}-\dot{T}H)f_{RT}
+\dot{T}^2\\\nonumber&\times&f_{RTT}\geqslant0,\\\nonumber
\textbf{WEC}&:& \rho+(\rho+p)f_T+\frac{1}{2}
f+3H^2(1-q)f_{R}+18H^4(j-q-2)f_{RR}\\\nonumber&-&3H\dot{T}f_{RT})\geqslant0,
\quad {\rho}_{eff}+p_{eff}\geqslant0,\\\nonumber \textbf{SEC}&:&
(\rho+3p)+(\rho+p)f_T-f-6H^2(1-q)f_{R}+3[6H^3(j-q-2)]^3\\\nonumber&\times&f_{RRR}-
18(s+j+q^2+7q+4)f_{RR}-36H^3(j-q-2)\dot{T}f_{RRT}\\\nonumber&+&3(\ddot{T}+
\dot{T}H)f_{RT}+3\dot{T}^2f_{RTT}\geqslant0, \quad
{\rho}_{eff}+p_{eff}\geqslant0,
\end{eqnarray}
\begin{eqnarray}\nonumber
\textbf{DEC}&:&
(\rho-p)+(\rho+p)f_T+f+6H^2(1-q)f_{R}-[6H^3(j-q-2)]^2\\\nonumber&\times&f_{RRR}
-6H^4(s+5j+(q-1)(q+4))f_{RR}+12H^3(j-q-2)\dot{T}\\\nonumber&\times&f_{RRT}
-(\ddot{T}+5\dot{T}H)f_{RT}-\dot{T}^2f_{RTT}\geqslant0,
{\rho}_{eff}+p_{eff}\geqslant0, \\\nonumber
&&{\rho}_{eff}\geqslant0.
\end{eqnarray}
The energy conditions in $f(R)$ gravity$^{{19})}$ can be recovered
for $f(R,T)=f(R)$ and also in case of GR for particular choice
$f(R,T)=R$. To illustrate how above conditions can be used to place
bounds on $f(R,T)$ gravity, we consider two particular forms of
$f(R,T)$ gravity,\\ \textbf{(i)} $f(R)+\lambda{T}$, \quad
\textbf{(ii)} $R+2f(T)$.

We shall obtain the power law solutions for each case and hence the
constraints set by the respective energy conditions.

\section{Power Law Solutions for $f(R,T)$ Gravity}

It is important to study the existence of exact power solutions
corresponding to different phases of cosmic evolution. Such
solutions are particularly relevant because in FRW background they
represent all possible cosmological evolutions such as radiation
dominated, matter dominated or dark energy eras. We discuss power
law solutions for two particular models of $f(R,T)$ gravity.

\subsection{$f(R,T)=f(R)+{\lambda}T$}

For the particular case  $f(R,T)=f(R)+{\lambda}T$$^{{11,13})}$, the
effective Einstein field equations are given by Eq.(\ref{6}) with
\begin{eqnarray}\nonumber
{T}_{{\mu}{\nu}}^{eff}&=&\frac{1}{f_{R}}\left[(1+\lambda)T_{\mu\nu}
+(\lambda{p}+\frac{1}{2}{\lambda}T)g_{\mu\nu}+\frac{1}{2}(f-Rf_{R})g_{\mu\nu}+
({\nabla}_{\mu}{\nabla}_{\nu}\right.\\\label{23}&-&\left.g_{{\mu}{\nu}}
{\Box})f_{R}\right],
\end{eqnarray}
where $f_R$ is the derivative of $f(R)$ with respect to scalar
curvature $R$. The Friedmann equation and the trace of the field
equations are given by
\begin{eqnarray}\label{24}
\mathbf{\Theta}^2=\frac{3}{f_R}\left[\rho+\lambda(\rho+p)+\frac{{\lambda}T}{2}
+\frac{1}{2}(f-Rf_R)-\mathbf{\Theta}\dot{R}f_{RR}\right],
\\\label{25} Rf_{R}+3{\Box}f_{R}(R,T)-2f=(1+3\lambda)T+4{\lambda}p,
\end{eqnarray}
where $\mathbf{\Theta}=3\dot{a}/a$ is the expansion scalar.

The standard matter satisfies the following energy conservation
equation
\begin{equation}\label{26}
\dot{\rho}=-\mathbf{\Theta}(\rho+p).
\end{equation}
For the homogeneous and isotropic spacetime, the field equations can
be represented by Raychaudhuri equation
\begin{eqnarray}\nonumber
\dot{\mathbf{\Theta}}+\frac{1}{3}{\mathbf{\Theta}}^2&=&-\frac{1}{2f_R}
\left[\rho+3p+4{\lambda}p-f+Rf_R+(3\ddot{R}+\mathbf{\Theta}\dot{R})f_{RR}
\right.\\\label{27}&+&\left.3\dot{R}^2f_{RRR}\right].
\end{eqnarray}
Combination of Raychaudhuri and Friedmann equations yields
\begin{equation}\label{28}
R=-2(\dot{\mathbf{\Theta}}+\frac{2}{3}\mathbf{\Theta}^2).
\end{equation}
We assume that there exists an exact power law solution to the
modified field equations
\begin{equation}\label{29}
a(t)=a_0t^m,
\end{equation}
where $m$ is a positive real number. If $0<m<1$, then the required
power law solution is decelerating while for $m>1$ it exhibits
accelerating behavior. For the equation of state $p=\omega{\rho}$,
the energy conservation equation leads to
\begin{equation}\label{30}
\rho(t)=\rho_0t^{-3m(1+\omega)}.
\end{equation}
Using Eq.(\ref{29}) in Eq.(\ref{28}), the scalar curvature becomes
\begin{equation}\label{31}
R=-6m(2m-1)t^{-2}=-\eta_{m}t^{-2},
\end{equation}
where $\eta_{m}=6m(2m-1)$. We see that sign of $R$ depends on the
value of $m$, $R>0$ if $0<m<\frac{1}{2}$ and $R<0$ for
$m>\frac{1}{2}$. Since $m=\frac{1}{2}$ leads to vanishing of $R$, so
we exclude this value of $m$ in our discussion.

Using Eqs.(\ref{30}) and (\ref{31}), Friedmann equation can be
written in terms of Ricci scalar $R$, $f$ and $f_R$ as
\begin{equation}\label{32}
f_{RR}R^2+\frac{m-1}{2}Rf_R+\frac{1-2m}{2}f-(2m-1)A\rho_0
\left(\frac{-R}{\eta_m}\right)^{\frac{3m(1+\omega)}{2}}=0,
\end{equation}
where $A=1+\frac{\lambda}{2}(3-\omega)$. This represents second
order differential equation for $f(R)$ whose general solution is
\begin{equation}\label{33}
f(R)=X_{m\omega}\left(\frac{-R}{\eta_m}\right)^{\frac{3m(1+\omega)}{2}}
+C_1R^{\frac{1}{4}(3-m-\sqrt{\delta_m})}+C_2R^{\frac{1}{4}(3-m+\sqrt{\delta_m})},
\end{equation}
where
\begin{equation*}
X_{m\omega}=\frac{4A(2m-1)\rho_0}{3m^2(3\omega+4)(\omega+1)-m(9\omega+13)+2},
\quad \delta_m=m^2+10m+1,
\end{equation*}
and $C_1,~C_2$ are arbitrary constants of integration. Since $m>0$,
so $\delta_m>0$ for cosmologically viable solutions. $X_{m\omega}$
is found to be real valued but it diverges for
$3m^2(3\omega+4)(\omega+1)-m(9\omega+13)+2=0$, \emph{i.e.}, $m$ and
$\omega$ satisfy any of the relations
$\omega=\frac{3-7m\pm\sqrt{\delta_m}}{6m}$ or
$m=\frac{13+9\omega\pm\sqrt{9\omega^2+66\omega+73}}{6(\omega+1)(3\omega+4)}$.
Since $R<0$, so $\left(-R/\eta_m\right)>0$ for all $R$, thus we have
real valued solution $f(R,T)=f(R)+\lambda{T}$ showing that the power
law solution exists for this model.

For $\lambda=0$, we obtain solution as in $f(R)$ gravity$^{{16})}$.
To check whether the $f(R,T)$ gravity reduces to GR, we need to put
$C_1=C_2=\lambda=0$. For $m=\frac{2}{3(1+\omega)}$ and
$\rho_0=\frac{4}{3(1+\omega)^2}$, this theory reduces to GR. We are
interested to construct the $f(R,T)$ model of the form
$\alpha{R}^n+\lambda{T}$. If we put $m=\frac{2n}{3(1+\omega)}$, then
$f(R)$ is given by
\begin{equation}\label{34}
f(R)=\alpha_{n\omega}(-R)^n,
\end{equation}
where
\begin{equation}\nonumber
\alpha_{n\omega}=\frac{2^{3-2n}3^{n-1}nA(n(4n-3(1+\omega))^{1-n}
(1+\omega)^{2n-2}}{(n^2(6\omega+8)-n(9\omega+13)+3(\omega+1))},
\end{equation}
and hence $f(R,T)=\alpha_{n\omega}(-R)^n+{\lambda}T$. This model
represents the exact Friedmann-like power law solution
$a\propto{t}^{\frac{2n}{3(1+\omega)}}$ and the limit $n\rightarrow1$
with $\lambda=0$ leads to GR. For $n=1$, our solutions represents
$\Lambda$CDM model of the form $f(R,T)=R+\lambda{T}$.

\subsubsection{Phantom Phase Power Law Solution}

We construct the phantom phase power law solution which lead to big
rip singularity. For this case, the scale factor and Hubble
parameter are expressed as
\begin{eqnarray}\nonumber
a(t)=a_0(t_s-t)^{-m}, \quad  H(t)=\frac{m}{t_s-t}.
\end{eqnarray}
The scale factor diverges within finite time $(t\rightarrow{t}_s)$
leading to big rip singularity for $m\geqslant1$$^{{29})}$. The
results for this case can be recovered by just replacing $m$ by $-m$
in the previous section. Hence, the phantom phase power law solution
exist for $f(R)+\lambda{T}$ gravity.

\subsubsection{Constraining $f(R,T)=f(R)+{\lambda}T$ Gravity}

In section \textbf{2}, we have found that Raychaudhuri equations
with attractive behavior of gravitational interaction give rise to
SEC and NEC which hold for any theory of gravitation. In this form
of $f(R,T)$ gravity, one can employ an approach similar to that in
GR to develop the energy conditions.

Equations (\ref{6})and (\ref{23}) can be written as
\begin{equation}\label{35}
R_{\mu\nu}=T_{\mu\nu}-\frac{T}{2}g_{\mu\nu},
\end{equation}
where
\begin{eqnarray}\nonumber
T_{\mu\nu}&=&\frac{1}{f_R}\left[(1+\lambda)T_{\mu\nu}+\frac{\lambda}{2}
(\rho-p)g_{\mu\nu}+({\nabla}_{\mu}{\nabla}_{\nu}-g_{{\mu}{\nu}}{\Box})f_{R}\right],\\\nonumber
T&=&\frac{1}{f_R}\left[(1+3\lambda)T+4\lambda{p}+f-Rf_R-3{\Box}f_{R}\right].
\end{eqnarray}
Equations (\ref{14}) and (\ref{35}) lead to following inequalities
\begin{eqnarray}\label{36}
\textbf{NEC}&:&\frac{1}{f_R}\left[(\rho+p)(1+\lambda)+(\ddot{R}-\dot{R}H)f_{RR}
+\dot{R}^2f_{RRR}\right]\geqslant0,\\\nonumber
\textbf{SEC}&:&\frac{1}{2f_R}\left[(\rho+3p)+4p\lambda-f+Rf_{R}
+3\dot{R}^2f_{RRR}+3(\ddot{R}\right.\\\label{37}&+&\left.\dot{R}H)f_{RR}\right]\geqslant0.
\end{eqnarray}
For $\lambda=0$, one can obtain the NEC and SEC in $f(R)$ gravity.
Furthermore, the more familiar forms of NEC and SEC in GR,
\emph{i.e.}, $\rho+p\geqslant0$ and $\rho+3p\geqslant0$, can be
recovered if $\lambda=0$ and $f(R)=R$.

To derive the WEC and DEC, we can extend the GR approach by
introducing an effective energy-momentum tensor. The above
inequalities of the SEC and NEC are obtained directly from
Raychaudhuri equations, however equivalent results can be derived by
using the conditions $\rho_{eff}+p_{eff}\geqslant0$ and
$\rho_{eff}+3p_{eff}\geqslant0$. From Eq.(\ref{23}), the effective
energy density and effective pressure are given by
\begin{eqnarray}\label{38}
{\rho}_{eff}&=&\frac{1}{f_R}\left[\rho+\frac{\lambda}{2}
(3\rho-p)+\frac{1}{2}(f-Rf_{R})-3H\dot{R}f_{RR}\right],\\\nonumber
{p}_{eff}&=&\frac{1}{f_R}\left[(p-\frac{\lambda}{2}
(\rho-3p)-\frac{1}{2}(f-Rf_{R})+(\ddot{R}+2\dot{R}H)f_{RR}
\right.\\\label{39}&+&\left.\dot{R}^2f_{RRR}\right].
\end{eqnarray}
The WEC and DEC in $f(R)+\lambda{T}$ gravity can be obtained by
following the effective energy-momentum tensor approach. The WEC is
obtained by satisfying inequality (\ref{36}) and the constraint
\begin{eqnarray}\label{40}
\frac{1}{f_R}\left[\rho+\frac{\lambda}{2}(3\rho-p)+\frac{1}{2}
(f-Rf_{R})-3H\dot{R}f_{RR}\right]\geqslant0,
\end{eqnarray}
and DEC is obtained by satisfying inequalities (\ref{36}),
(\ref{40}) and the constraint
\begin{eqnarray}\label{41}
\frac{1}{f_R}\left[(\rho-p)+2\lambda(\rho+p)+f-Rf_{R}-
(\ddot{R}+5\dot{R}H)f_{RR}-\dot{R}^2f_{RRR}\right]\geqslant0.
\end{eqnarray}
We find that by setting $\lambda=0$, the WEC and DEC in $f(R)$
gravity can be recovered. Moreover, for $f(R)=R$ and $\lambda=0$,
the well-known form of weak and dominant energy conditions in GR can
be reproduced.

The above energy conditions can be used to put constraints on a
given $f(R)$ model in the context of $f(R,T)$ gravity.  We assume
that $f_R>0$ to keep the effective gravitational constant positive.
Using the relations (\ref{21}), the energy conditions for
$f(R)+\lambda{T}$ gravity in terms of present day values of
$H,~q,~j$ and $s$ are given by
\begin{eqnarray}\nonumber
\textbf{NEC}&:& (1+\lambda)(\rho_0+p_0)-6H^4(s_0-j_0
+(q_0+1)(q_0+8))f_{0RR}+H_0^4[6\\\nonumber&\times&H_0(j_0-q_0-2)]^2
f_{0RRR}\geqslant0,\\\nonumber
\textbf{WEC}&:&\rho_0+\frac{\lambda}{2}(3\rho_0-p_0)+\frac{1}{2}
f_0+3H_0^2(1-q_0)f_{0R}+18H_0^4[j_0-q_0-2]\\\nonumber&\times&f_{0RR}\geqslant0,
\quad {\rho}_{eff}+p_{eff}\geqslant0,\\\nonumber \textbf{SEC}&:&
(\rho_0+3p_0)+4p_0\lambda-f_0-6H_0^2(1-q_0)f_{0R}+3[6H_0^3(j_0-q_0-2)]^2
\\\nonumber&\times&f_{0RRR}-18H_0^4[s_0+j_0+q_0^2+7q_0+4]f_{0RR}
\geqslant0, \quad{\rho}_{eff}+p_{eff}\geqslant0,\\\nonumber
\textbf{DEC}&:&
(\rho_0-p_0)+2\lambda(\rho_0-p_0)+f_0+6H_0^2(1-q_0)f_{0R}-[6H_0^3(j_0-q_0
\\\nonumber&-&2)]^2f_{0RRR}-6H^4[s_0+5j_0+(q_0-1)(q_0+4)]f_{0RR}\geqslant0, \quad
\\\nonumber &&{\rho}_{eff}+p_{eff}\geqslant0, \quad
{\rho}_{eff}\geqslant0.
\end{eqnarray}
In order to present the concrete application of the above energy
conditions in $f(R,T)$ gravity, we employ the exact power law
solution of $f(R)+\lambda{T}$ gravity. We consider the present day
values of deceleration, jerk and snap parameters as
$q_0=-0.81\pm0.14,~j_0=2.16^{+0.81}_{-0.75}$ and
$s_0=-0.22^{+0.21}_{-0.19}$$^{{28})}$. We shall discuss the WEC
requirement to illustrate how the above conditions can be used to
place constraints on $f(R,T)$ gravity. We note that all the above
conditions depend on the present value of pressure $p_0$, so for
simplicity we assume $p=0$.

Now, we take the power law solution as an objective model which is
given by
\begin{equation}\label{42}
f(R,T)=\alpha_{n}(-R)^n+{\lambda}T,
\end{equation}
where $n$ is an integer and
$\alpha_n=\frac{2^{3-2n}3^{n-1}nA(4n^2-3n)^{1-n}}{(8n^2-13n+3)}$.
The constraints to fulfill the WEC, \emph{i.e.},
$\rho_{eff}\geqslant0$, $\rho_{eff}+p_{eff}\geqslant0$, are
respectively obtained as
\begin{eqnarray}\label{43}
&&(2+3\lambda)\rho_0+\alpha_n[6H^2_0(1-q_0)]^n[B_1(n^2-n)-n+1]\geqslant0,
\\\nonumber
&&(1+\lambda)\rho_0+\alpha_nn(n-1)6H^4_0[6H^2_0(1-q_0)]^{n-2}[-(s_0-j_0
+(q_0+1)\\\label{44}&\times&(q_0+8))-B_2(n-2)]\geqslant0,
\end{eqnarray}
where $B_1=(j_0-q_0-2)/(1-q_0)^2$ and $B_2=(j_0-q_0-2)^2/(1-q_0)$.
As the standard matter is assumed to satisfy the necessary energy
conditions and $\lambda>0$, so $(2+3\lambda)\rho_0>0$ and
$(1+\lambda)\rho_0>0$. Hence, the inequality (\ref{43}) is reduced
to
\begin{equation}\nonumber
\alpha_n(3.3H_0)^{2n}\beta_n\geqslant0, \quad \texttt{where} \quad
\beta_n=B_1(n^2-n)-n+1.
\end{equation}
It is clear from above expression, the result is trivial for
$n=0,1$. We consider the following two cases:\\
(i) $\alpha_n>0$, the allowed values for $n$ are $n=\{2,3,4,...\}$.
Now $\beta_n>0$ in the range $n=\{4,5,6,...\}$ and $\beta_n<0$ for
$n=2,3$.\\
(ii) $\alpha_n<0$, the acceptable values of $n$ are
$n=\{-1,-2,...\}$ and in this particular range we have $\beta_n<0$.
Thus, the inequality $\rho_{eff}\geqslant0$ is satisfied for
$n=\{...,-2,-1,4,5,...\}$.

Now we check the validity of Eq.(\ref{44}) except $n=0,1$ as the
result is trivial for this choice. The inequality is transformed to
the following form
\begin{equation}\nonumber
\alpha_n(3.3H_0)^{2n-2}\mu_n\geqslant0, \quad \texttt{where} \quad
\mu_n=(n^2-n)(2.054-0.52n).
\end{equation}
The results of the above inequality can be interpreted as:\\
(i) $\mu_n>0$, if $n=\{2,3,-1,-2,...\}$ and for $\mu_n<0$, the
acceptable values of $n$ are $n=\{4,5,6,...\}$.\\
(ii) $\alpha_n>0$, with acceptable range $n=\{2,3,4,...\}$ and
$\alpha_n<0$, when $n=\{-1,-2,-3,...\}$. Hence, the condition
$\rho_{eff}+p_{eff}\geqslant0$ is satisfied for $n=2,3$.

\subsection{$f(R,T)=R+2f(T)$}

Now, we construct the power law solutions for $R+2f(T)$ gravity,
where $f(T)$ is an arbitrary function of the trace of
energy-momentum tensor. The effective Einstein field equations are
given by Eq.(\ref{6}) with
\begin{equation}\nonumber
{T}_{{\mu}{\nu}}^{eff}=(1+2f_T)T_{\mu\nu}+(2pf_T+f)g_{\mu\nu},
\end{equation}
where $f_T$ is the derivative of $f$ with respect to the trace of
energy-momentum tensor $T$. The Friedmann equation and the trace
equation can be obtained as
\begin{equation}\label{46}
\mathbf{\Theta}^2=3[\rho+2(\rho+p)f_T+f], \quad
R=-(\rho-3p)-2(\rho+p)f_T-4f.
\end{equation}
It follows that the field equations can be represented as the
Raychaudhuri equation
\begin{equation}\label{47}
\dot{\mathbf{\Theta}}+\frac{1}{3}{\mathbf{\Theta}}^2=-\frac{1}{2}
\left[(\rho+3p)+2(\rho+p)f_T-2f\right].
\end{equation}

Combining Eqs.(\ref{46}) and (\ref{47}), we can get the Ricci scalar
$R$ given in Eq.(\ref{28}). Using Eq.(\ref{30}), the Friedmann
equation can be written in terms of trace of the energy-momentum
tensor $T,~f(T)$ and its derivative with respect to $T$ as
\begin{equation}\label{48}
Tf_T+\frac{T}{2(1+\omega)}+\frac{(1-3\omega)f}{2(1+\omega)}-
\frac{K(1-3\omega)T^{\frac{2}{3m(1+\omega)}}}{2(1+\omega)}=0,
\end{equation}
where $K=3m^2(\rho_0(1-3\omega))^{\frac{-2}{3m(1+\omega)}}$. This is
the first order differential equation in $f(T)$ whose solution is
\begin{equation}\label{49}
f(T)=\frac{T}{\omega-3}+L_{m\omega}T^{\frac{2}{3m(1+\omega)}}+
C_1T^{\frac{-(1-3\omega)}{2(1+\omega)}},
\end{equation}
where
$L_{m\omega}=\frac{9m^3(1-3\omega)(\rho_0(1-3\omega))^{\frac{-2}
{3m(1+\omega)}}}{4+3m(1-3\omega)}$ and $C_1$ is arbitrary constant
of integration, $L_{m\omega}$ is finite and real valued unless
$4+3m(1-3\omega)=0$. In general, the function $f(T)$ is real valued
if $m$ and $\omega$ do not satisfy the relation
$m=\frac{-4}{3(1-3\omega)}$ and if $\omega>3$. Therefore, the power
law solutions exist for $R+2f(T)$ gravity.

For $m=0$, we have $a=a_0$, so that $H=R=0$, it represents the
Einstein static universe. The solution for this case is
\begin{equation}\label{50}
f(T)=\frac{T}{\omega-3}+C_1T^{\frac{-(1-3\omega)}{2(1+\omega)}}.
\end{equation}
The standard Einstein gravity can be recovered for the choice
$C_1=0$, $m=\frac{2}{3(1+\omega)}$ and
$\rho_0=\frac{4}{3(1+\omega)^2}$. In order to develop a more general
form of function $f(T)$, we put $m=\frac{2n}{3(1+\omega)}$, so that
\begin{equation}\label{51}
f(T)=\frac{T}{\omega-3}+a_{n\omega}T^n,
\end{equation}
where $a_{n\omega}=\frac{2^{3-2n}3^{n-1}n^{3-2n}(1-3\omega)^{1-n}
(1+\omega)^{2n-2}}{4(1+\omega)+2n(1-3\omega)}$. We can ensure that
this theory reduces to GR for $n=1$. It is remarked that phantom
power law solutions exist for this form of $f(R,T)$ gravity, which
can be obtained in a similar fashion as in section \textbf{4.1.1}.

\subsubsection{Constraining $f(R,T)=R+2f(T)$ Gravity}

The effective energy density $\rho_{eff}$ and effective pressure
$p_{eff}$ for this particular $f(R,T)$ gravity are defined as
\begin{equation}\label{52}
\rho_{eff}=\rho+2(\rho+p)f_T+f, \quad  p_{eff}=p-f.
\end{equation}
Using Eq.(\ref{52}) in energy conditions (\ref{17})-(\ref{20}), the
following form is obtained$^{{26})}$
\begin{eqnarray}\nonumber
\textbf{NEC}&:& (\rho+p)[1+2f_T]\geqslant0,\\\nonumber
\textbf{WEC}&:& \rho+2(\rho+p)f_T+f\geqslant0, \quad
{\rho}_{eff}+p_{eff}\geqslant0,\\\nonumber \textbf{SEC}&:&
\rho+3p+2(\rho+p)f_T-2f\geqslant0,
\quad{\rho}_{eff}+p_{eff}\geqslant0,\\\nonumber \textbf{DEC}&:&
\rho-p+2(\rho+p)f_T+2f\geqslant0, \quad {\rho}_{eff}\geqslant0,
\quad {\rho}_{eff}+p_{eff}\geqslant0.
\end{eqnarray}
To check how these conditions place bounds on power law solution
(\ref{51}) in $R+{\lambda}T$ gravity we put $p=0$, so that $T=\rho$.
Hence, the function $f(\rho)$ is of the form
\begin{equation}\label{53}
f(\rho)=-\frac{\rho}{3}+a_{n}{\rho}^n,
\end{equation}
where $a_{n}=\frac{2^{2(1-n)}3^{n-1}n^{3-2n}}{n+2}$. The constraints
to accomplish the above energy conditions are obtained as follows:
\begin{eqnarray}\nonumber
\textbf{NEC}&:& \frac{\rho}{3}+2na_n{\rho}^n\geqslant0,\\\nonumber
\textbf{WEC}&:& \frac{\rho}{3}+5na_n{\rho}^n\geqslant0, \\\nonumber
\textbf{SEC}&:& \frac{4\rho}{3}+2(2n+1)a_n{\rho}^n\geqslant0,
\\\nonumber \textbf{DEC}&:& \rho+(7n+2)a_n{\rho}^n\geqslant0.
\end{eqnarray}
Above conditions are trivially satisfied for $n=0,1$. The quantities
$na_n$, $(2n+1)a_n$ and $(7n+2)a_n$ are negative when
$n=\{-3,-4,-5,...\}$ and positive for $n=\{-1,2,3,...\}$. Since
$\rho$ is assumed to be positive, so it is obvious that these
conditions are satisfied within the range of $n=\{-1,2,3,...\}$.

\section{Stability of Power Law Solutions}

In this section, we are interested to study the stability of power
law solutions against linear perturbations in $f(R,T)$ gravity.
First, we assume a general solution $H(t)=H_h(t)$ for the
cosmological background of FRW universe that satisfies
Eqs.(\ref{24}) and (\ref{46}). The matter fluid is assumed to be
dust and evolution of the matter energy density can be expressed in
terms of $H_h(t)$ as
\begin{equation}\label{54}
\rho_h(t)=\rho_0 e^{-3\int{H_h(t)dt}},
\end{equation}
where $\rho_0$ is an integration constant. Since the matter
perturbations also contribute to the stability, so we introduce
perturbations in Hubble parameter and energy density to study the
perturbation around the arbitrary solution $H_h(t)$ as
follows$^{{30})}$
\begin{equation}\label{55}
H(t)=H_h(t)(1+\delta(t)), \quad \quad \rho(t)=\rho_h
(1+\delta_m(t)).
\end{equation}
In the following, we develop perturbation equations for two specific
cases $f(R,T)=f(R)+{\lambda}T$ and $f(R,T)=R+2f(T)$.

\subsection{$f(R,T)=f(R)+{\lambda}T$}

To study the linear perturbations, we expand function $f(R)$ in
powers of $R_h$ evaluated at $H(t)=H_h(t)$ as
\begin{equation}\label{56}
f(R)=f^h+f^h_R(R-R_h)+\mathcal{O}^2,
\end{equation}
where function $f(R)$ and its derivative are evaluated at $R_h$. The
term $\mathcal{O}^2$ includes all the terms proportional to the
square or higher powers of $R$. The Ricci scalar $R$ at
$H(t)=H_h(t)$ is given by
\begin{equation}\label{57}
R_h=-6(\dot{H_h}+2H_h^2).
\end{equation}
By introducing the expressions (\ref{55}) and (\ref{56}) in the FRW
equation (\ref{24}), the equation for the perturbation $\delta(t)$
becomes
\begin{equation}\label{58}
\dot{\delta}(t)+c(t)\delta(t)=\frac{A\rho_h}{3H_hR_hf^h_{RR}}\delta_m,
\end{equation}
where
\begin{eqnarray*}
c(t)=\frac{d}{dt}\left[ln\left(H_h^{-1}R_h^2f_{RR}^h\right)\right]+H_h
[2(\frac{d}{dt}[ln(f_R^h)])^{-1}-1].
\end{eqnarray*}
The conservation equation (\ref{26}) implies the second perturbation
equation as
\begin{equation}\label{59}
\dot{\delta}_m(t)+3H_h(t)\delta(t)=0.
\end{equation}
We can eliminate $\delta(t)$ from Eqs.(\ref{58}) and (\ref{59}) and
arrive at the following second-order perturbation equation
\begin{equation}\label{60}
\ddot{\delta}_m(t)+c_1(t)\dot{\delta}_m(t)+\frac{A\rho_h}{3H_hR_hf^h_{RR}}\delta_m=0,
\end{equation}
where
\begin{eqnarray*}
c_1(t)=\frac{d}{dt}\left[ln\left(H_h^{-2}R_h^2f_{RR}^h\right)\right]+H_h
[2(\frac{d}{dt}[ln(f_R^h)])^{-1}-1].
\end{eqnarray*}

Here, we consider the $f(R,T)$ model proposed in section
\textbf{4.1} for the dust case which is defined as
$f(R)=\alpha_{n}(-R)^n+{\lambda}T$. We evaluate $f(R)$ and its
derivatives at $H(t)=H_h(t)$ and hence the perturbation
$\delta_m(t)$ is given by
\begin{equation}\label{61}
\delta_m(t)=C_{+}t^{\mu_+}+C_{-}t^{\mu_-},
\end{equation}
where $C_{\pm}$ are arbitrary constants and
\begin{eqnarray*}
\mu_{\pm}=\frac{8n^2-15n+13}{6(n-1)}\pm\frac{\sqrt{n^2(8n^2-15n+3)^2+18
(8n^3-21n^2+16n-3)\rho_0}}{6n(n-1)}.
\end{eqnarray*}
In order to study the stability of perturbation given by
Eq.(\ref{61}), one needs to check the signs of exponents
$\mu_{\pm}$. The exponents are found to be negative provided that
$n\leqslant-2$, otherwise $\mu_\pm$ would be positive and the
perturbation is unstable. The perturbation $\delta(t)$ is found to
be
\begin{equation}\label{62}
\delta(t)=\frac{-1}{3H_h}(C_{+}\mu_+t^{\nu_+}+C_{-}\mu_-t^{\nu_-}),
\end{equation}
where $\nu_{\pm}=\mu_{\pm}-1$. It can be seen that exponent $\nu_+$
is negative for $n\leqslant-2$ and $\nu_-$ is always negative.
Hence, as the time evolves the condition $n\leqslant-2$ ensures the
decay of perturbations $\delta(t)$ and $\delta_m(t)$ which implies
the stability of power law solution for this $f(R,T)$ gravity.

\subsection{$f(R,T)=R+2f(T)$}

We explore the behavior of perturbations (\ref{55}) for this
$f(R,T)$ model and expand the function in powers of $T_h(=\rho_h$)
as$^{{26})}$
\begin{equation}\label{63}
f(T)=f^h+f^h_T(T-T_h)+\mathcal{O}^2,
\end{equation}
where $\mathcal{O}$ term includes all the terms proportional to the
squares or higher powers of $T$. The function $f(T)$ and its
derivatives are evaluated at $T=T_h$. Using Eqs.(\ref{55}) and
(\ref{63}) in FRW equation (\ref{46}), it follows that
\begin{equation}\label{64}
(T_h+3T_hf^h_T+2T^2_hf_{TT}^h)\delta_m(t)=6H_h^2\delta(t).
\end{equation}
Combining Eqs.(\ref{59}) and (\ref{64}), the first order matter
perturbation equation is
\begin{equation}\label{65}
\dot{\delta}_m(t)+\frac{1}{2H_h}(T_h+3T_hf_T^h+2T_h^2f_{TT}^h)\delta_m(t)=0.
\end{equation}
which leads to
\begin{equation}\label{66}
\delta_m(t)=C_4\exp\left\{\frac{-1}{2}\int{C_T}dt\right\}, \quad
C_T=\frac{T_h}{H_h}(1+3f_T^h+2T_hf_{TT}^h).
\end{equation}
The behavior of perturbation $\delta(t)$ can be seen from the
relation
\begin{equation}\label{67}
\delta(t)=\frac{C_4C_T}{6H_h}\exp\left\{\frac{-1}{2}\int{C_T}dt\right\}.
\end{equation}

We explore the stability of power law model (proposed in section
\textbf{4.2}) of the form
\begin{equation}\label{68}
f(T)=a_1T+a_2T^n,
\end{equation}
where $a_1$ and $a_2$ are parameters. One can evaluate the
expression $C_T$ and integral $\frac{-1}{2}\int{C_Tdt}$ for the
model (\ref{66}) as
\begin{eqnarray}\label{69}
C_T=\frac{3}{2n}\left[\rho_0(3a_1+1)t^{-2n+1}+a_2\rho_0^nn(2n+1)t^{-2n^2+1}\right],\\\label{70}
\frac{-1}{2}\int{C_Tdt}=\frac{3}{8n(n-1)}\left[\rho_0(3a_1+1)t^{-2(n-1)}
+\frac{a_2\rho_0^nn(2n+1)}{n+1}t^{-2(n^2-1)}\right].
\end{eqnarray}
As the time evolves, we need to set the conditions for decay of
perturbations. It is obvious that expression (\ref{69}) and
(\ref{70}) decay as time increases for the choice $n>1$ which
results in decay of $\delta(t)$ and $\delta_m(t)$. Hence, for large
values of $t$ perturbation decays, this corresponds to the stability
of power law solutions for $R+2f(T)$ gravity. We find that the
conditions developed for stability are compatible with some
constraints to fulfil the energy conditions. Hence, we may remark
that power law solutions are acceptable regarding to the stability,
energy conditions and late time acceleration of the universe.

\section{Conclusions}

The issue of accelerated expansion of the universe can be explained
by taking into account the modified theories of gravity such as
$f(R,T)$ gravity. The $f(R,T)$ gravity provides an alternative way
to explain the current cosmic acceleration with no need of
introducing either the existence of extra spatial dimension or an
exotic component of DE. In this modified gravity, cosmic
acceleration may result not only due to geometrical contribution to
the total cosmic energy density but it also depends on matter
contents. This theory depends upon matter source term, so each
choice of matter Lagrangian $\mathcal{L}_m$ would generate a
specific set of field equations. The various forms of Lagrangian in
this gravity give rise to question how to constrain the $f(R,T)$
gravity theories on physical grounds. We have made an attempt to
address this issue and classify the particular $f(R,T)$ models. In
this respect, we have developed the energy conditions on general as
well as particular forms of this gravity. The energy conditions in
modified theories of gravity have a well defined physical
motivation, \emph{i.e.,} Raychaudhuri's equation along with
attractive nature of gravity.

The energy conditions and exact power law solutions are studied in
the framework of $f(R,T)$ gravity. We have derived the energy
conditions directly from effective energy-momentum tensor approach
under the transformation $\rho{\rightarrow}\rho_{eff}$ and
$p{\rightarrow}p_{eff}$. The general inequalities imposed by these
conditions are presented in terms of deceleration $(q)$, jerk $(j)$
and snap $(s)$ parameters. In order to get some insights on the
application of these conditions, we consider two particular forms of
$f(R,T)$ gravity, \emph{i.e.}, $f(R,T)=f(R)+{\lambda}T$ and
$f(R,T)=R+2f(T)$. In standard paradigm, the expansion history of the
universe underwent a power law decelerating phase followed by late
time acceleration. Therefore, power law solutions are important in
cosmology to represent the matter dominated phase that later
connects to an accelerating phase. We have shown that exact power
law solutions exist for a special class of $f(R,T)$ models. These
solutions mimic the $\Lambda$CDM model as particular case. We have
obtained the necessary constraints to fulfil the energy conditions
for this particular class of $f(R,T)$ gravity. We summarize the
results of these two models as follows: {\begin{itemize}
\item {$f(R)+{\lambda}T$}
\end{itemize}}
It is shown that exact power law solution exists for this form of
$f(R,T)$ gravity given in Eq.(\ref{33}). In the limit of
$\lambda=0$, the corresponding result can be recovered in $f(R)$
gravity. To ensure that this theory reduces to GR, we need to set
$C_1=C_2=\lambda=0$, with $m=\frac{2}{3(1+\omega)}$ and
$\rho_0=\frac{4}{3(1+\omega)^2}$. We have constructed the general
form of $f(R)+{\lambda}T$ model which corresponds to $R^n$ gravity.
For this particular model of $f(R,T)$ gravity, the NEC and SEC are
derived from the Raychaudhuri equation together with the condition
that gravity is attractive. It is shown that these conditions differ
from those derived in the context of GR and $f(R)$
gravity$^{{19})}$. The general expression of weak and dominant
energy conditions are obtained by introducing the effective
energy-momentum tensor in the context of GR. We have examined the
WEC bounds on $\alpha_{n}(-R)^n+{\lambda}T$ model in terms of
present day observational values $H_0,~q_0,~j_0$ and $s_0$.
{\begin{itemize}
\item {$R+2f(T)$}
\end{itemize}}
The model $f(R,T)=R+2f(T)$ corresponds to gravitational Lagrangian
with time dependent cosmological constant being function of trace of
the energy-momentum tensor$^{{31})}$. This model appears to be
interesting and has widely been studied in literature$^{{9}-{12})}$.
A general form of $f(T)$ model (\ref{51}) is obtained which
corresponds to GR in the limit $n=1$. We have applied the energy
conditions to set the possible constraints on this $f(R,T)$ model.
It is found that energy conditions are globally satisfied within the
range of $n=\{-1,2,3,...\}$. It is worth mentioning here that
results of power law solutions and energy conditions obtained in
this paper are quite general which correspond to GR and $f(R)$
gravity.

We have also analyzed the stability of power law solutions under
linear homogeneous perturbations in the FRW background for $f(R,T)$
gravity. In particular, perturbations for energy density and Hubble
parameter are introduced which produce linearized perturbed field
equations. It is shown that stability/instability can be studied for
particular $f(R,T)$ models under some restrictions. The stability
conditions are found to be compatible with energy conditions bounds
to some extent. Hence, power law solutions in $f(R,T)$ gravity can
be considered as viable models to explain the cosmic history of the
universe. It is also interesting to note that for these $f(R,T)$
models, the gravitational coupling becomes an effective and time
dependent coupling which modifies the gravitational interaction
between matter and curvature. Our analysis shows that stable power
law solutions are contained in class of $f(R,T)$ models, at least
considering a given background evolution of the universe.\\

{\bf Acknowledgment}

\vspace{0.25cm}

The authors would like to thank the Higher Education Commission,
Islamabad, Pakistan for its financial support through the
\emph{Indigenous Ph.D. 5000 Fellowship Program Batch-VII}.\\\\
1) C. L. Bennett, M. Halpern, G. Hinshaw, N. Jarosik, A. Kogut, M.
Limon, S. S. Meyer, L. Page, D. N. Spergel, G. S. Tucker, E.
Wollack, E. L. Wright, C. Barnes, M. R. Greason, R. S. Hill, E.
Komatsu, M. R. Nolta, N. Odegard, H. V. Peiris, L. Verde and J. L.
Weiland: Astrophys. J. Suppl. \textbf{148} (2003) 1; D. N. Spergel,
L. Verde, H. V. Peiris, E. Komatsu, M. R. Nolta, C. L. Bennett, M.
Halpern, G. Hinshaw, N. Jarosik, A. Kogut, M. Limon, S. S. Meyer, L.
Page, G. S. Tucker, J. L. Weiland, E. Wollack and E. L. Wright:
Astrophys. J. Suppl. \textbf{148} (2003) 175; D. N. Spergel, R.
Bean, O. Doré, M. R. Nolta, C. L. Bennett, J. Dunkley, G. Hinshaw,
N. Jarosik, E. Komatsu, L. Page, H. V. Peiris, L. Verde, M. Halpern,
R. S. Hill, A. Kogut, M. Limon, S. S. Meyer, N. Odegard, G. S.
Tucker, J. L. Weiland, E. Wollack and E. L. Wright:
Astrophys. J. Suppl. \textbf{170} (2007) 377.\\
2) S. Perlmutter, S. Gabi, G. Goldhaber, A. Goobar, D. E. Groom, I.
M. Hook, A. G. Kim, M. Y. Kim, G. C. Lee, R. Pain, C. R.
Pennypacker, I. A. Small, R. S. Ellis, R. G. McMahon, B. J. Boyle,
P. S. Bunclark, D. Carter, M. J. Irwin, K. Glazebrook, H. J. M.
Newberg, A. V. Filippenko, T. Matheson, M. Dopita and W. C. Couch:
Astrophys. J. \textbf{483} (1997) 565; S. Perlmutter, G. Aldering,
M. D. Valle, S. Deustua, R. S. Ellis, S. Fabbro, A. Fruchter, G.
Goldhaber, A. Goobar, D. E. Groom, I. M. Hook, A. G. Kim, M. Y. Kim,
R. A. Knop, C. Lidman, R. G. McMahon, P. Nugent, R. Pain, N.
Panagia, C. R. Pennypacker, P. Ruiz-Lapuente, B. Schaefer and N.
Walton: Nature \textbf{391} (1998) 51; S. Perlmutter, G. Aldering,
G. Goldhaber, R. A. Knop, P. Nugent, P. G. Castro, S. Deustua, S.
Fabbro, A. Goobar, D. E. Groom, I. M. Hook, A. G. Kim, M. Y. Kim, J.
C. Lee, N. J. Nunes, R. Pain, C. R. Pennypacker, R. Quimbey, C.
Lidman, R. S. Ellis, M. Irwin, R. G. Mcmahon, P. Ruiz-lapuente, N.
Walton, B. Schaefer, B. J. Boyle, A. V. Filippenko, T. Matheson, A.
S. Fruchter, N. Panagia, H. J. M. Newberg and W. J. Couch:
Astrophys. J. \textbf{517} (1999) 565; A. G. Riess, L. G. Strolger,
J. Tonry, Z. Tsvetanov, S. Casertano, H. C. Ferguson, B. Mobasher,
P. Challis, N. Panagia, A. V. Filippenko, W. Li, R. Chornock, R. P.
Kirshner, B. Leibundgut, M. Dickinson, A. Koekemoer, N. A. Grogin
and M. Giavalisco : Astrophys. J. \textbf{607} (2004) 665; A. G.
Riess, L. G. Strolger, S. Casertano, H. C. Ferguson, B. Mobasher, B.
Gold, P. J. Challis, A. V. Filippenko, S. Jha, W. Li, J. Tonry, R.
Foley, R. P. Kirshner, M. Dickinson, E. MacDonald, D. Eisenstein, M.
Livio, J. Younger, C. Xu, T. Dahlén and D. Stern: Astrophys. J.
\textbf{659} (2007) 98.\\
3) E. Hawkins, S. Maddox, S. Cole, O.
Lahav, D. S. Madgwick, P. Norberg, J. A. Peacock, I. K. Baldry, C.
M. Baugh, J. Bland-Hawthorn, T. Bridges, R. Cannon, M. Colless, C.
Collins, W. Couch, G. Dalton, R. D. Propris, S. P. Driver, S.P., G.
Efstathiou, R. S. Ellis, C.S. Frenk, K. Glazebrook, C. Jackson, B.
Jones, I. Lewis, S. Lumsden, W. Percival, B. A. Peterson, W.
Sutherland and K. Taylor: Mon. Not. Roy. Astron. Soc. \textbf{346}
(2003) 78; M. Tegmark, M. A. Strauss, M. R. Blanton, K. Abazajian,
S. Dodelson, H. Sandvik, X. Wang, D. H. Weinberg, I. Zehavi, N. A.
Bahcall, F. Hoyle, D. Schlegel, R. Scoccimarro, M. S. Vogeley, A.
Berlind, T. Budavari, A. Connolly, D. J. Eisenstein, D. Finkbeiner,
J. A. Frieman, J. E. Gunn, L. Hui, B. Jain, D. Johnston, S. Kent, H.
Lin, R. Nakajima, R. C. Nichol, J. P. Ostriker, A. Pope, R.
Scranton, U. Seljak, R. K. Sheth, A. Stebbins, A. S. Szalay, I.
Szapudi, Y. Xu, J. Annis, J. Brinkmann, S. Burles, F. J. Castander,
I. Csabai, J. Loveday, M. Doi, M. Fukugita, B. Gillespie, G.
Hennessy, D. W. Hogg, Z. E. Ivezic´, G. R. Knapp, D. Q. Lamb, B. C.
Lee, R. H. Lupton, T. A. McKay, P. Kunszt, J. A. Munn, L. Connell,
J. Peoples, J. R. Pier, M. Richmond, C. Rockosi, D. P. Schneider, C.
Stoughton, D. L. Tucker, D. E. V. Berk, B. Yanny and D. G. York:
Phys. Rev. D \textbf{69} (2004) 103501; S. Cole , W. J. Percival, J.
A. Peacock, P. Norberg, C. M. Baugh, C. S. Frenk, I. Baldry, J. B.
Hawthorn, T. Bridges, R. Cannon, M. Colless, C. Collins, W. Couch,
N. J. G. Cross, G. Dalton, V. R. Eke, R. D. Propris, S. P. Driver,
G. Efstathiou, R. S. Ellis, K. Glazebrook, C. Jackson, A. Jenkins,
O. Lahav, I. Lewis, S. Lumsden, S. Maddox, D. Madgwick, B. A.
Peterson, W. Sutherland and K. Taylor: Mon. Not. Roy.
Astron. Soc. \textbf{362} (2005) 505.\\
4) D. J. Eisentein,  I. Zehavi, D. W. Hogg, R. Scoccimarro, M. R.
Blanton, R. C. Nichol, R. Scranton, Hee-Jong Seo, M. Tegmark, Z.
Zheng, S. F. Anderson, J. Annis, N. Bahcall, J. Brinkmann, S.
Burles, F. J. Castander, A. Connolly, I. Csabai, M. Doi, M.
Fukugita, J. A. Frieman, K. Glazebrook, J. E. Gunn, J. S. Hendry, G.
Hennessy, Z. Ivezic, S. Kent, G. R. Knapp, H. Lin, Yeong-Shang Loh,
R. H. Lupton, B. Margon, T. A. McKay, A. Meiksin, J. A. Munn, A.
Pope, M. W. Richmond, D. Schlegel, D. P. Schneider, K. Shimasaku, C.
Stoughton, M. A. Strauss, M. SubbaRao, A. S. Szalay, I. Szapudi, D.
L. Tucker, B. Yanny and D. G. York: Astrophys. J. \textbf{633}
(2005) 560.\\
5) B. Jain and A. Taylor: Phys. Rev. Lett. \textbf{91} (2003)
141302.\\
6) S. Nojiri and S. D. Odintsov: Int. J. Geom. Methods Mod. Phys.
\textbf{4} (2007) 115; T. P. Sotiriou and S. Liberati: Ann. Phys.
\textbf{322} (2007) 935; T. P. Sotiriou, and V. Faraoni: Rev. Mod.
Phys. \textbf{82} (2010) 451; A. De Felice and S. Tsujikawa: Living
Rev. Rel. \textbf{13} (2010) 3; S. Nojiri and S. D. Odintsov: Phys.
Rep. \textbf{505} (2011) 59. K. Bamba, S. Capozziello, S. Nojiri and
S. D. Odintsov: Astrophys. Space Sci. \textbf{345} (2012) 155.\\
7) R. Ferraro and F. Fiorini: Phys. Rev. D \textbf{75} (2007) 08403;
G. R. Bengochea and R. Ferraro: Phys. Rev. D \textbf{79}
(2009)124019; E. V. Linder: Phys. Rev. D \textbf{81} (2010) 127301.
K. Bamba, C. Q. Geng, C. C. Lee: arXiv:1008.4036v1; K. Bamba, C. Q.
Geng, C. C. Lee and L. W. Luoa: JCAP \textbf{01}
(2011) 021.\\
8) S. M. Carroll, A. De Felice, V. Duvvuri, D. A. Easson, M. Trodden
and M. S. Turner: Phys. Rev. D \textbf{71} (2005) 063513; G.
Cognola, E. Elizalde, S. Nojiri, S. D. Odintsov and S. Zerbini:
Phys. Rev. D \textbf{73} (2006) 084007; S. Nojiri and S. D.
Odintsov: Prog. Theor. Phys. Suppl. \textbf{172} (2008)
81.\\
9) T. Harko, F. S. N. Lobo, S. Nojiri and S. D. Odintsov: Phys. Rev.
D \textbf{84} (2011) 024020.\\
10) M. J. S. Houndjo and O. F. Piattella: Int. J. Mod. Phys. D
\textbf{21} (2012) 1250024.\\
11) M. J. S. Houndjo: Int. J. Mod. Phys. D \textbf{21} (2012)
1250003.\\
12) M. Jamil, D. Momeni, M. Raza and R. Myrzakulov: Eur. Phys. J. C
\textbf{72} (2012) 1999.\\
13) M. Sharif and M. Zubair: J. Phys. Soc. Jpn. \textbf{81} (2012) 114005.\\
14) M. Sharif and M. Zubair: J. Cosmology Astropart. Phys.
\textbf{03} (2012) 028 [Errata \textbf{05} (2012) E01].
\emph{Thermodynamic Behavior of $f(R,T)$ Gravity Models at the
Apparent Horizon} (To appear in Cent. Eur. J.
Phys).\\
15) N. Goheer, R. Goswami, P. K. S. Dunsby and K. Ananda: Phys. Rev.
D \textbf{79} (2009) 121304; A. R. Rastkar, M. R. Setare and F.
Darabi: Astrophys. Space Sci. \textbf{337} (2012) 478; M. R. Setare
and F. Darabi: Gen. Relativ. Gravit. \textbf{44} (2012) 2521.\\
16) N. Goheer, J. Larena and P. K. S. Dunsby: Phys. Rev. D
\textbf{80} (2009) 061301.\\
17)S. W. Hawking and G. F. R. Ellis: \emph{The Large Scale Structure
of Spacetime}, (Cambridge University Press, England, 1973).\\
18) M. Visser: Phys. Rev. D \textbf{56} (1997) 7578; J. Santos and
J. S. Alcaniz: Phys. Lett. B \textbf{642} (2006) 311; J. Santos, J.
S. Alcaniz and M. J. Reboucas: Phys. Rev. D \textbf{74} (2006)
067301; Y. Gong, A. Wang, Q. Wu and Y.-Z. Zhang: JCAP \textbf{08}
(2007)
018; Y. Gong and A. Wang: Phys. Lett. B \textbf{652} (2007) 63.\\
19) J. Santos, J. S. Alcaniz, M. J. Reboucas and F. C. Carvalho:
Phys. Rev. D \textbf{78} (2007) 083513; J. Santos, M. J. Reboucas
and J. S. Alcaniz: Int. J. Mod. Phys. D \textbf{19} (2010) 1315.\\
20) J. Wang, Y.-B. Wua, Y.-X. Guob, W.-Q. Yang and L. Wanga:
Phys. Lett. B \textbf{689} (2010) 133.\\
21) O. Bertolami and M. C. Sequeira: Phys. Rev. D \textbf{79} (2009)
104010.\\
22) N. M. Garcia: Phys. Rev. D \textbf{83} (2011) 104032.\\
23) Y. Y. Zhao: Eur. Phys. J. C \textbf{72} (2012) 1924.\\
24) K. Atazadeh, A. Khaleghi, H. R. Sepangi, Y. Tavakoli: Int. J.
Mod. Phys. D \textbf{18} (2009) 1101.\\
25) D. Liu and M. J. Reboucas: Phys. Rev. D \textbf{86} (2012) 083515.\\
26) A. G. Alvarenga, M. J. S. Houndjo, A. V. Monwanou and J. B. C.
Orou: arXiv:1205.4678v2.\\
27) L. D. Landau and E. M. Lifshitz: \emph{The Classical Theory of
Fields} (Butterworth-Heinemann, 2002).\\
28) T. Chiba and T. Nakamura: Prog. Theor. Phys. \textbf{100} (1998)
1077; V. Sahni, T. D. Saini, A. A. Starobinsky and U. Alam: JETP
Lett. \textbf{77} (2003) 201; Pisma Zh. Eksp. Teor. Fiz. \textbf{77}
(2003) 249; M. Visser: Class. Quantum Grav. \textbf{21} (2004) 2603;
D. Rapetti, S. W. Allen, M. A. Amin and R. D. Blandford: Mon. Not.
R. Astron. Soc. \textbf{375} (2007) 1510; N. J. Poplawski: Class.
Quantum Grav. \textbf{24} (2007) 3013.\\
29) S. Nojiri, S. D. Odintsov and S. Tsujikawa: Phys. Rev. D
\textbf{71} (2005) 063004; K. Bamba, R. Myrzakulov, S. Nojiri and S.
D. Odintsov: Phys. Rev. D \textbf{85} (2012) 104036.\\
30) A. de la Cruz-Dombriz and D.
Sáez-Gómez: arxiv:1112.4481\\
31) N. J. Poplawski: arXiv:gr-qc/0608031.\\
\end{document}